\documentstyle[12pt]{article}  
  \oddsidemargin=\evensidemargin
 \let\msk=\medskip 
  
 \let\qqd=\qquad  \def\ve{\vfil\eject}

   \let\d=\delta \let\e=\varepsilon
    
    \let\p=\pi \let\r=\rho
   \let\c=\chi

\def\0{\over }    \def\1{\vec }   \def\2{{1\over2}} \def\3{{\ss}}
\def\4{{1\over4}} \def\5{\bar }   \def\6{\partial } \def\7#1{{#1}\llap{/}}
\def\8#1{{\textstyle{#1}}}        \def\9#1{{\bf {#1}}}
\def\_#1{$\underline{\hbox{#1}}$} \def\^#1{$\overline{\hbox{#1}}$}

\def\<{\langle } \def\>{\rangle }  
 
\def \({\left( } \def \){\right) }

  \let\ex=\times   
      \let\and=\wedge
     
\def\|#1{{}_{\bigg|_{#1}}}

\ifx\Box  \else \fi

\def\pmbf#1{\setbox0=\hbox{${#1}$}   \kern-.025em\copy0\kern-\wd0
      \kern.05em\copy0\kern-\wd0     \kern-.025em\raise.0433em\box0 }

\def\inbar{\vrule height1.5ex width.7pt depth0pt} 
\def\ifundefined#1{\expandafter\ifx\csname#1\endcsname\relax}
\makeatletter \ifundefined{new@mathgroup} {} \else \input{oldlfont.sty} \fi
\def\ZZ{\relax{\sf Z\kern-.4em \sf Z}}  \def\IR{\relax{\rm I\kern-.18em R}}
\def\IN{\relax{\rm I\kern-.18em N}} \def\IP{\relax{\rm I\kern-.18em P}}
\def\IQ{\relax\,\hbox{$\inbar\kern-.3em{\rm Q}$}}
\def\IC{\hbox{\,$\inbar\kern-.3em{\sf C}$}}
\def\citen#1{\if@filesw \immediate\write \@auxout {\string\citation{#1}}\fi%
\@tempcntb\m@ne \let\@h@ld\relax \def\@citea{}%
\@for \@citeb:=#1\do {\@ifundefined {b@\@citeb}%
    {\@h@ld\@citea\@tempcntb\m@ne{\bf ?}%
    \@warning {Citation `\@citeb ' on page \thepage \space undefined}}%
    {\@tempcnta\@tempcntb \advance\@tempcnta\@ne
    \setbox\z@\hbox\bgroup\ifcat0\csname b@\@citeb \endcsname \relax
       \egroup \@tempcntb\number\csname b@\@citeb \endcsname \relax
       \else \egroup \@tempcntb\m@ne \fi \ifnum\@tempcnta=\@tempcntb
       \ifx\@h@ld\relax \edef \@h@ld{\@citea\csname b@\@citeb\endcsname}%
       \else \edef\@h@ld{\hbox{--}\penalty\@highpenalty
	      \csname b@\@citeb\endcsname}\fi
    \else \@h@ld\@citea\csname b@\@citeb \endcsname \let\@h@ld\relax \fi}%
 \def\@citea{,\penalty\@highpenalty\hskip.13em plus.13em minus.13em}}\@h@ld}
\def\@citex[#1]#2{\@cite{\citen{#2}}{#1}}%
\def\@cite#1#2{\leavevmode\unskip
  \ifnum\lastpenalty=\z@\penalty\@highpenalty\fi
  \ [{\multiply\@highpenalty 3 #1
  \if@tempswa,\penalty\@highpenalty\ #2\fi}]}   
\makeatother 

\def\beq{\begin{equation}} \def\eeq{\end{equation}} \def\eql#1{\label{#1}\eeq}
\def\bea{\begin{eqnarray}} \def\eea{\end{eqnarray}} 
\def\fnote#1#2{\begingroup\def\thefootnote{#1}\footnote{#2}
	   \addtocounter{footnote}{-1}\endgroup}    

\def\plb#1 #2 {Phys. Lett. {\bf B#1} #2 }
\def\phr#1 #2 {Phys. Rep. {\bf  #1} #2 } 
\def\npb#1 #2 {Nucl. Phys. {\bf B#1} #2 }
\def\aph#1 #2 {Ann. Phys. {\bf #1} #2 }  
\def\jmp#1 #2 {J. Math. Phys. {\bf #1} #2 }
\def\prd#1 #2 {Phys. Rev. {\bf D#1} #2 }
\def\prl#1 #2 {Phys. Rev. Lett. {\bf #1} #2 }
\def\rmp#1 #2 {Rev. Mod. Phys.  {\bf #1} #2 }
\def\zpc#1 #2 {Z. Phys. {\bf #1C} #2 }
\def\cmp#1 #2 {Commun. Math. Phys. {\bf #1} #2 }
\def\mpl#1 #2 {Mod. Phys. Lett. {\bf A#1} #2 }
\def\ijmp#1 #2 {Int. J. Mod. Phys. {\bf A#1} #2 }

\let\3=\ss \catcode`\"=\active \let"=\"
\ifundefined{emptypage} \let\emptypage=\relax \fi

\long\def\del#1\enddel{ }
\def\LG{Landau--Ginzburg} \def\CY{Calabi--Yau} \def\MS{mirror symmetry}
 \def\[{\left[} \def\]{\right]} \def\cR{{\cal R}}

\def\ng{n_{27}} \def\na{n_{\overline{27}}} 
\def\config#1 {\ifnum #1=4 \configv \else \configf \fi}
\def\configv#1 #2 #3 #4 #5 {\hbox{$\IC_{(#1,#2,#3,#4)}[#5]$}}
\def\configf#1 #2 #3 #4 #5 #6 {\hbox{$\IC_{(#1,#2,#3,#4,#5)}[#6]$}}
\def\spec#1 #2 #3 {$\ng=#1$, $\na=#2$ $\c=#3$: }
\def\cfa{\rlap{\IC$^B$}\config } \def\cfb{\rlap{\IC$^A$}\config }
\def\cI#1)#2]{$\IC_{#1)}^I#2]$}     \def\cF#1)#2]{$\IC_{#1)}^F#2]$}
\def\cB#1)#2]{$\IC_{#1)}^{D}#2]$} \def\cC#1)#2]{$\IC_{#1)}^{I,C}#2]$}
\def\cD#1)#2]{$\IC_{#1)}^{I,D}#2]$} \def\cE#1)#2]{$\IC_{#1)}^{I,C\cap D}#2]$}

\def\VS#1#2{\vrule height #1mm depth #2mm width 0pt}
\def\TS{@{~~\VS{5}0}} \def\TL{\VS02}        

\def\figuresonly{\pagestyle{empty}\figa\ve\figb\ve\figc\end{document}}
\long\def\old#1\endold{{\small #1}}         \def\oldansw{o } \def\cutansw{c }
\newcount{\npre} \npre=1  \def\negansw{s } \def\figansw{f } \def\textansw{t }
\def\ifpre{\ifnum\npre=1 } \def\ifsub{\ifnum\npre=0 }        \def\cut#1{#1}
\def\askversion{\message{
Preprint (p) / submit (s) / text only (t) / figures only (f):  (p/s/t/f)? }
    \read-1 to\answ \ifx\answ\negansw \npre=0 \else \npre=1 \fi
    \ifx\answ\figansw { } \else \def\figuresonly{ }   \fi
    \ifx\answ\oldansw \def\old##1\endold{{\small ##1}}\fi
    \ifx\answ\textansw \npre=2 \else \message{
Cut figures (use 'c' in case of memory problem):  (c/n)? }
    \read-1 to\answ\ifx\answ\cutansw \def\cut##1{}\npre=7\fi\fi \figuresonly }

\def\bpic{\begin{picture}} \def\epic{\end{picture}} \thicklines
\def\lab#1)#2#3{\put#1){\makebox(0,0)[#2]{\small #3}}}
\def\putlin#1,#2,#3,#4,#5){\put#1,#2){\line(#3,#4){#5}}} 
\def\putvec#1,#2,#3,#4,#5){\put#1,#2){\vector(#3,#4){#5}}}

\newcount{\vmul} \newcount{\hdiv} 
\newcount{\wi}   \newcount{\he}   
\newcount{\hq}   \newcount{\vq}   
\newcount{\hoff} \newcount{\auxc} \newcounter{figco}   \def\npt{\circle*{2}}

\def\vlline{\put(-3,0){\line(1,0)6}}      

\def\putvm#1{\mbox{\bpic(0,0)\funit=1pt\vlline\epic}}   

\def\hlabo{\put(0,0){\mbox{\bpic(0,0)\funit=1pt\put(0,-3){\line(0,1)3}\epic}}}

\def\Vpt#1,#2){\hq=#2\advance\hq by -#1 \multiply\hq by 2 \divide\hq by\hdiv
	       \vq=#1\advance\vq by #2\multiply\vq by\vmul\put(\hq,\vq){\npt}}
\def\Vplo#1{\vbox{\hdiv=2\vmul=1 \figsca \auxc=\he \multiply\auxc by\vmul
    \hoff=\wi\divide\hoff by2\stepcounter{figco}\message{[Fig. \arabic{figco}}
    \begin{center}\let\.=\Vpt \bpic(\wi,\auxc)(-\hoff,0) \figlab #1 \hlabo
    \put(-\hoff,0){\framebox(\wi,\auxc){}} \epic \\[5mm]
    Fig. \arabic{figco}: \figcap \end{center}} \vfil \message{]}}
\def\figsca{\unitlength=1.1pt \wi=500 \he=400} \let\funit=\unitlength

\begin{document}     \def\cern{CERN-TH.6802/93}
{\hfill\cern\vskip-9pt   \hfill hep-th/9303015}
\vskip 25mm \centerline{\hss\LARGE\bf   Where are the Mirror Manifolds? \hss}
\begin{center} \vskip 18mm
       Maximilian KREUZER\fnote{*}{e-mail: kreuzer@cernvm.cern.ch} 
\vskip 5mm
       CERN, Theory Division\\
       CH--1211 Geneva 23, SWITZERLAND

\vfil                        {\bf ABSTRACT}                \end{center}

The recent classification of Landau--Ginzburg potentials and their abelian
symmetries focuses attention on a number of models with large positive Euler
number for which no mirror partner is known. All of these models are related
to Calabi--Yau manifolds in weighted $\IP_4$, with a characteristic structure
of the defining polynomials. A closer look at these potentials suggests a
series of non-linear transformations, which relate the models to configurations
for which a construction of the mirror is known, though only at certain
points in moduli space. A special case of these transformations generalizes
the $\ZZ_2$ orbifold representation of the $D$ invariant, implying a
hidden symmetry in tensor products of minimal models.

\vfil\noindent \cern\\ February 1993 \msk
  \thispagestyle{empty} \newpage  \emptypage
  \setcounter{page}{1} \pagestyle{plain}

\section{Introduction}

Mirror symmetry~\cite{gp,cls,co}, which relates string compactifications
with exchanged numbers of $27$ and $\overline{27}$ representations of $E_6$,
provides a powerful guiding principle in striving for
completeness in the classification of string vacua.
In $N=2$ superconformal field theory this symmetry is realized by the natural
operation of redefining the right-moving $U(1)$ charge~\cite{d}.
For various explicit constructions, as for example \CY\
manifolds~\cite{cy}, on the other hand, the existence of
such a mirror partner is a highly non-trivial matter.

A large class of string vacua with $(2,2)$ superconformal invariance can be
constructed from \LG\ models~\cite{mvw,lvw}, which are particularly useful
in case of $N=2$ supersymmetry because of a non-renormalization theorem for
the superpotential.
Recently the non-degenerate potentials that give rise to $N=2$ superconformal
models with central charge $c=9$, as needed for the internal sector, and all
their abelian symmetries have been classified by
construction~\cite{cqf,nms,ks,aas}.
This requires consideration of up to 9 chiral superfields; the models with up
to 5 fields, which already represent 70\% of all configurations, are related
to \CY\ manifolds in weighted $\IP_4$ with the same defining
polynomial~\cite{gvw,w}.

For a subclass of the non-degenerate potentials a construction of the mirror
model was given by Berglund and H"ubsch~\cite{bh}. Their method was fully
confirmed, for the relevant class of polynomials, by the calculation of
all abelian \LG\ orbifolds~\cite{aas}. The full set of models, however,
features a striking lack of \MS: 810 of the 3837 orbifold spectra have no
mirror. In fact, this problem was apparent already in the
untwisted case~\cite{nms}. The new spectra from orbifolding all have Euler
numbers $\chi\le 480$, whereas there is a remarkable set of potentials
with large positive Euler numbers, for which no mirror model is known yet.
These ``singlet spectra'' range up to
$\chi=840$, already close to the maximal value of $\chi=960$.

It is this set of polynomials that we want to analyse in the present paper.
All of them have 4 or 5 non-trivial fields, and thus define \CY\
manifolds. As there is only a handful of models with, say, $\chi>500$,
we expect that it is easier in this class
to find some structure that may be relevant for
the lack of \MS. Indeed, 7 of the 9 singlet spectra in this range
come from a polynomial of class VI in Arnold's classification~\cite{agv}; the
remaining two require only a slight modification, with couplings among four
fields. Moreover, each spectrum can be obtained from a number of different
configurations and orbifolds, which, however, look very similar. It is thus
easy to find non-linear transformations with constant determinant~\cite{gvw},
which indicate that they are, in fact, in (almost) all cases equivalent.

In section~2 we consider the model with $\chi=840$ in some detail.
Here we already find the essential structures and the non-linear
transformations relating different configurations.
In this case, they can even be used to represent the model in a Fermat
configuration, so that for a deformation of an orbifold we would
know how to construct the mirror.

In section~3 we then list all the singlet
models along decreasing (modulus of the) Euler number until we hit, at
$|\chi|=450$, the first singlet spectrum with negative $\c$.
Most of these models exactly follow the pattern found in section~2.
Only at $\c=540$ do we find a new type of polynomial. Furthermore, this
spectrum apparently comes from two inequivalent configurations. A systematic
search for a non-linear relation, instead, reveals a non-linear symmetry,
present in each of the two models. Nevertheless, a connection between
the two configurations, involving deformations, orbifolding, and a mirror map,
can be found.

In section~4 we generalize these non-linear transformations
to arbitrary non-degenerate polynomials and give
a simple algorithm for checking the conditions that have to be fulfilled by
the exponents. Then we briefly discuss the implied hidden symmetries
in the special case of tensor products of
minimal models. In this case the transformation has already been found
in~\cite{ls}. Sections~3 and 4 are almost independent and could be exchanged.
Section~5 contains our conclusions.

\section{Cutting loops and trees}

In this section we analyse the model with Euler number 840. 
But first we need to introduce some notation.

A configuration $\IC_{(n_1,\ldots,n_I)}[d]$ denotes the set of non-degenerate
polynomials that are quasi-homogeneous of degree $d$ in the
superfields $X_i$ with respect
to weights $n_i$. In weighted $\IP_4$, $I$ must be 5 and the condition for
vanishing first Chern class is $\sum n_i=d$. For \LG\ models we can have an
arbitrary number $I$ of fields, but the central charge $c=3\sum(1-2n_i/d)$
is required to be 9 for the internal sector of a heterotic
string. This coincides with the \CY\ condition for $I=5$; for $I=4$ we have
to add a trivial (massive) field with $n_i=d/2$ to make contact with \CY\
manifolds. For convenience, however, we will often omit such fields in the
following; their appropriate transformation under symmetry groups to make
determinants positive should be understood implicitly.

In both frameworks we need to require that the quasi-homogeneous polynomial
$W(X_i)$ is non-degenerate (or transversal), i.e. that the origin is the only
place where all gradients vanish.
This implies, in particular, that for each field $X_i$ there
must be a monomial of the form $X_i^{a_i}X_j$~\cite{agv} (the coefficients
can be normalized to 1). We call the sum of these $I$ terms the skeleton of
$W$, and say that $X_i$ points at $X_j$ if $i\neq j$.
If there is more than one pointer with the same target, then
non-degeneracy requires the presence of additional monomials, which we call
links (the details will not be important in the following and can be found in
ref.~\cite{cqf}). Note that the skeleton
already determines the configuration. A given configuration, on the other
hand, in general admits a number of inequivalent skeletons.

The Berglund--H"ubsch construction of the mirror manifold~\cite{bh} now
applies exactly if no links are required, i.e. if the polynomial is equal to
its skeleton. We call such a polynomial invertible; the mirror can then be
constructed as a particular orbifold with the exponents $a_i+\d_{ij}$
transposed along each chain of pointers in the skeleton.

For constructing a heterotic string with space-time supersymmetry we need to
project the \LG\ model to integer charges, i.e. mod out the 
symmetry
$\ZZ_d(n_1,\ldots,n_I)$, whose generator acts by multiplication with a phase
$\exp(2\p i\,n_i/d)$ on the field $X_i$~\cite{iv}. This projection will always
be assumed and it is only in case of additional twists that we will use the
term orbifold.
The gauge non-singlet particle spectrum at low energies is then determined by
the numbers $\ng=b_{12}$ and $\na=b_{11}$ of chiral primary fields with
charges $(Q_L,Q_R)=(1,1)$ and $(1,2)$, respectively~\cite{g}, where $b_{ij}$
are the Hodge numbers of the corresponding \CY\ manifold, if it exists.

The spectra with the largest values of the Euler number $\c=2(\na-\ng)$ found
in refs.~\cite{nms,ks} are $(\ng,\na;\c)$=(11,491;960), (12,462;900),
(13,433;840), (14,416;804), \ldots, of which already the third 
one does not have a known mirror partner. This singlet spectrum comes from
the three configurations 
{\config 5 24 31 244 567 866 1732 },
{\config 5 36 31 366 866 1299 2598 }, and
{\config 5 18 31 183 634 433 1299 },
each of which has a unique skeleton:
\del
\beq X_1^{62}X_3+X_2^{48}X_3+X_3^7X_1+X_4^3X_2+X_5^2\;+\;\e X_1^{67-31k}
     X_2^{4+24k}, \label f \eeq
\beq X_1^{62}X_3+X_2^{72}X_3+X_3^7X_1+X_4^3+X_5^2\;+\;\e X_1^{67-31k}
     X_2^{6+36k}, \label s \eeq
\beq X_1^{62}X_3+X_2^{36}X_3+X_3^7X_1+X_4^2X_2+X_5^3\;+\;\e X_1^{67-31k}
     X_2^{3+18k}. \label l \eeq
\enddel
\bea &X_1^{62}X_3+X_2^{48}X_3+X_3^7X_1+X_4^3X_2+X_5^2\;+\;\e X_1^{67-31k}
     X_2^{4+24k},& \label f \\
     &X_1^{62}X_3+X_2^{72}X_3+X_3^7X_1+X_4^3+X_5^2\;+\;\e X_1^{67-31k}
     X_2^{6+36k},& \label s \\
     &X_1^{62}X_3+X_2^{36}X_3+X_3^7X_1+X_4^2X_2+X_5^3\;+\;\e X_1^{67-31k}
     X_2^{3+18k}.& \label l \eea
Here we have added the simplest link monomials that make the polynomials
non-degenerate 
(the number of monomials of degree $d$ is 28 in all three cases).
A graphical representation of the potentials, with a dot for each field, an
arrow for each pointer, and a dashed line for the link, is given in fig.~1A
for the polynomials (\ref f) and (\ref l), and in fig.~1B for the second
case.

\begin{figure} \begin{center} \funit=1.5pt                      
\newsavebox{\figa} \newsavebox{\figb} \newsavebox{\figc} \newsavebox{\link}
\def\dot#1){\put#1){\circle*3}}
\def\negtri#1{\multiput(0.05,-0.1)(0.1,-0.2){#1}{{\funit=1pt \circle*{1}}}}
\savebox\link(0,0)[tl]{\bpic(0.9,1.8)\negtri8\epic}
\savebox\figc(60,20)[bl]{\bpic(60,20)
   \putlin(0,0,1,1,20) \putvec(10,10,-1,-1,2)
   \putlin(0,0,1,0,30) \putvec(15,0,-1,0,3)
   \multiput(20.7,21.7)(1.8,-3.6){6}{\usebox\link} 
   \lab(20,25)b1\dot(20,20) \lab(30,-5)t2\dot(30,0) \lab(0,-5)t3\dot(0,0)
   \lab(55,-5)t4\dot(55,0)  \lab(55,25)b5\dot(55,20)            \epic}
\savebox\figb(60,20)[bl]{\bpic(60,20) \put(20,0){\oval(40,40)[tl]}
   \putvec(5.6,18,1,1,0)   \epic} 
\savebox\figa(60,20)[bl]{\bpic(60,20) \putlin(30,0,1,0,25)
   \putvec(40,0,-1,0,0)  \epic}
\bpic(255,35)\put(0,15){\usebox\figa}\put(0,15){\usebox\figb}
       \put(0,15){\usebox\figc} \lab(28,0)t{(A)}
   \put(100,15){\usebox\figb} \put(100,15){\usebox\figc} \lab(128,0)t{(B)}
   \put(200,15){\usebox\figc} \lab(228,0)t{(C)}
\epic\\[9mm]
Fig. 1: Graphical representation of the model with $\c=840$.
\end{center}\end{figure}

The structure and the exponents of the polynomials (\ref f)--(\ref l) almost
coincide, suggesting that the respective conformal field theories might be
identical. Greene, Vafa and Warner~\cite{gvw} have argued that \LG\ models
should flow to equivalent renormalization group fixed points if the potentials
can be related by a change of variables with constant determinant, provided
that multiple coverings are taken into account by appropriate orbifolding.
Indeed, such a map is easily found:
\beq    X_2\to X_2^{n\0n-1},\qqd X_4\to X_4 X_2^{-1\0n-1},          \eql z
transforms (\ref f) and (\ref l) into (\ref s), where $n=3$ and
$n=2$, respectively.
The effect on the skeletons is to ``cut'' the pointer from $X_4$ to $X_2$ and
to change the exponent of the target field of that pointer.
The transformation has constant determinant for arbitrary
$n$ and is $n-1$ to $n$. These multiplicities, however, are automatically
taken care of by the canonical $\ZZ_d$ twist, as $d=4p$, $6p$ and $3p$,
respectively, where $p=433$ is the $86^{th}$ prime number.

Having seen that all three representations of the model with $\c=840$ are
related by non-linear transformations, with the correct multiplicities
accidentally provided by the ratios of the degrees, it is natural to look for
orbifold representations of this model. Of course, the fact that abelian
orbifolds did not provide any {\it new} spectra with $\c>480$~\cite{aas} does
not exclude this possibility. Indeed, there is even an orbifold realization
in the {\it Fermat} configuration $\IC_{(1,1,12,28)}[84]$, which has the
spectrum (11,491;960) with the largest value of the Euler number.
We cannot find $\chi=840$ for an invertible skeleton in that configuration,
though, because then we would know the mirror. So we have to start from
the polynomial
\beq Y_1^{72}Y_3+Y_2^{72}Y_3+Y_3^7+Y_4^3+\e Y_1^{78-36k}Y_2^{6+36k} \eql2
and mod out the product of the 
groups $\ZZ_2(0,1,0,0)$, $\ZZ_3(1,1,0,1)$, $\ZZ_7(3,3,1,0)$, $\ZZ_8(1,7,0,0)$,
and $\ZZ_9(1,8,0,0)$.
Here the generator $g_{84}$ of the canonical $\ZZ_d$ is given by the product
$g_{84}=g_2g_3g_7(g_8)^2$ of the respective generators.
\footnote{The configuration $\IC_{(1,1,12,28)}[84]$ also accommodates the
transpose of $X_1^{84}X_2+X_2^{83}+X_3^7+X_4^3$. This  poly\-nomial
belongs to the configuration with the maximal degree $d=3486$ and the minimal
Euler number $\c=-960$.}

Again, there is a striking similarity between the polynomials (\ref 2) and
(\ref s), which suggests to us to look for a non-linear relation.
A straightforward calculation shows that
\beq X_1=Y_1^{504\0433},\qqd X_2=Y_2Y_1^{1\0433},\qqd X_3=Y_3Y_1^{-72\0433},
     \qqd X_4=Y_4 \eql c
does the job and has constant determinant. This transformation is 
433 to 504, so that the $\ZZ_{d}$ orbifold of (\ref s) is indeed mapped
onto the above $\ZZ_{504}\ex\ZZ_6$ orbifold of $(\ref2)$. Here the effect
of the transformation is to cut the pointer from $X_3$ to $X_1$, i.e. to
cut the loop, as is seen in fig.~1C, the graphical representation of
(\ref 2).

The construction of the mirror manifold for a Fermat polynomial is well
established~\cite{gp}, so for a deformation of an orbifold of our model we
would know how to proceed
(the untwisted model can be considered as an orbifold with
respect to the quantum symmetry~\cite{v} of the orbifold).
The trouble is, however, that we do not know
how to mod a quantum symmetry of a \CY\ manifold. In the \LG\ framework,
on the other hand, where these symmetries
usually are accessible by discrete torsion~\cite{fk},
it is not clear how to deform the model by moduli that are not
polynomial deformations but come from twisted sectors.

Note that these non-linear relations have non-trivial
implications for the underlying conformal field theories.
If the different orbifold models indeed flow to the same conformal field
theory, then that theory should have all of the respective quantum
symmetries. Unfortunately, however, in each of our \LG\ formulations,
only part of that full symmetry would be manifest.

\section{More missing mirror models}

\begin{table} \centering
\caption{Singlet spectra with the skeleton of fig. 1.} \vspace{7 mm}
\begin{tabular}{||c|c\TS c|l|l||l|c|c||} \hline\hline \TL 
$\c$& $\ng$& $\na$ & Configuration& Exponents& Orbifold& Twist&$b_1$\\\hline
840 & 13 & 433 &
\cfb4 24 31 244 567 1732    &~62 48 7 3  &\cI(2,3,24,55)[168]&433/504&72\\&&&
\cfa4 36 31 366 866 2598    &~62 72 7 3  &\cF(1,1,12,28)[84] &433/504&72\\&&&
\cfb5 18 31 183 634 433 1299 &~62 36 7 2 3&\cI(1,2,12,41,28)[84]&433/504&72\\
648 & 17 & 341 &
\cfb4 24 19 444 191 1356   &~38 48 3 7  &\cI(6,7,168,71)[504]&113/168&56\\&&&
\cfa4 28 19 518 226 1582   &~38 56 3 7  &\cF(1,1,28,12)[84]  &113/168&56\\&&&
\cfb5 14 19 259 386 113 791 &~38 28 3 2 7&\cI(1,2,28,41,12)[84]&113/168&56\\
612 & 20 & 326 &
\cfb4 12 31 184 423 1300  &~93 36 7 3  &\cI(1,3,18,41)[126]&325/378&108\\ &&&
\cfa4 18 31 276 650 1950  &~93 54 7 3  &\cF(1,2,18,42)[126]&325/378&108\\ &&&
\cfb5 9 31 138 472 325 975 &~93 27 7 2 3&\cI(1,4,18,61,42)[126]&325/378&108\\
576 & 7 & 295 &
\cfb4 28 37 144 381 1180     & ~37 28 8 3  &\cI(2,3,12,31)[96]&295/336&42\\&&&
\cfa4 42 37 216 590 1770     & ~37 42 8 3  &\cF(1,1,6,16)[48]&295/336&42\\&&&
\cfb5 21 37 108 424 295 885  & ~37 21 8 2 3&\cI(1,2,6,23,16)[48]&295/336&42\\
528 & 27 & 291 &
\cfb4 8 31 164 375 1156   &124 32 7 3  &\cI(2,9,48,109)[336]&289/336&144\\&&&
\cfa4 12 31 246 578 1734  &124 48 7 3  &\cF(1,3,24,56)[168] &289/336&144\\&&&
\cfb5 6 31 123 418 289 867 &124 24 7 2 3&\cI(1,6,24,81,56)[168]&289/336&144\\
516 & 36 & 294 &
\cfa4 7 247 41 590 1770    &247 7 43 3&\cI(1,36,6,86)[258]&295/301&252\\
510 & 38 & 293 &
\cfa5 4 494 41 343 147 1029 &247 2 25 3 7&\cI(2,252,21,175,75)[525]&49/50&252
\\
468 & 36 & 270 &
\cfb4 6 31 154 351 1084  & 155 30 7 3 &\cI(1,6,30,68)[210]&271/315&180\\&&&
\cfa4 9 31 231 542 1626  & 155 45 7 3 &\cI(1,4,30,70)[210]&271/315&180\\
456 & 37 & 265 &
\cfb4 148 7 24 351 1060     & 7 148 38 3   & ~~~~~~~ ------ & --- & -- \\&&&
\cfa4 222 7 36 530 1590     & 7 222 38 3   & ~~~~~~~ ------ & --- & -- \\&&&
\cfb5 111 7 18 394 265 795  & 7 111 38 2 3 & ~~~~~~~ ------ & --- & -- \TL\\
\hline\hline \end{tabular} \end{table} 

In this section we list the singlet models with $\chi\ge450$ along
decreasing Euler number. All of the corresponding 30 configurations with
13 different spectra only admit a unique skeleton. Furthermore, all of
these skeletons contain a loop and for 9 of the spectra the structure of
the polynomial is the one shown in fig.~1.
These 9 models are listed in table~1: For each spectrum the respective
configurations are printed with a superscript (A) or (B), indicating which
of the graphs in fig.~1 applies. Then we list the
exponents of the fields in the skeleton. In all cases with more than one
configuration the different polynomials are related by the
transformation~(\ref z).

As for the model with $\c=840$ we are interested in constructing an
equivalent orbifold representation in an invertible configuration (i.e.
a configuration admitting an invertible skeleton).
In most cases this can be done by using a transformation like (\ref c) to
cut the pointer from $X_3$ to $X_1$. Note that it is not
possible to cut one of the pointers at $X_3$, because then we would arrive
at an invertible skeleton and would know how to construct the mirror.
The last 3 columns of table~1 show the results of this computation:
First we give the configuration of the orbifold, with the superscript
indicating if it is invertible or of the Fermat type. The ratio $m/n 
$ in the column labelled ``Twist'' means that the map is $m$ to $n$, so that
$nd/m$ is the order of the twist group in the orbifold representation (which
must be a multiple of the degree of that configuration);
$b_1$ is the new exponent of $X_1$ and the graph of the skeleton is shown
in fig.~1C, where an additional pointer from $X_4$ to $X_2$ should be
supplemented if we start from fig.~1A (for more details on these
transformations see section~4 below).

Unfortunately, the lower right corner of table~1 is empty: The spectrum with
$\c=456$ is, in fact, the only case where we do not find any possibility
to cut the loop, let alone a transformation to an invertible configuration.
A search of the results of ref.~\cite{aas} for orbifold representations of
the spectrum confirms that indeed such a transformation does not exist.

\begin{table} \centering
\caption{Singlet spectra with the skeleton of fig. 2.} \vspace{7 mm}
\begin{tabular}{||c|c\TS c|l|l||l|c|c||} \hline\hline \TL 
$\c$& $\ng$& $\na$ & Configuration& Exponents& Orbifold& Twist&$b_I$\\ \hline
 540 & 14 & 284 &
\cfb5 19 11 112 153 276 571 &22 38 5 3 2&\cC(2,1,12,16,29)[60]&571/660&44\\&&&
\cfa4 19 22 224 306 1142    &44 38 5 3  &\cC(1,1,12,16)[60]   &571/660&44\\
 540 & 14 & 284 &
\cfb5 19 13 186 77 276 571  &26 38 3 5 2&\cC(2,1,20,8,29)[60] &571/780&52\\&&&
\cfa4 19 26 372 154 1142    &52 38 3 5  &\cC(1,1,20,8)[60]    &571/780&52\\
 512 & 15 & 271 &
\cfa4 147 15 29 353 1088   & 5 ~49 37 3  &\cC(10,1,2,24)[74]&544/555&50\\
      &&&&&\cB(196,20,29,490)[1470]&544/735&50\\
      &&&&&\cE(20,2,3,50)[150]&272/375&50\\
 476 & 16 & 254 &
\cfa4 15 138 26 332 1022   &46 ~5 34 3 &\cD(10,92,13,230)[690]&511/690&46\\&&&
\cfb5 15 69 13 166 248 511 &23 ~5 34 3 2&\cB(20,92,13,230,335)[690]&511/690&46
\\
 450 & 39 & 264 &
\cfa5 382 4 113 31 265 795 & 2 191 7 22 3&\cC(111,1,33,9,77)[231]&265/308&222
\TL\\ \hline\hline \end{tabular} \end{table} 
\begin{figure}
\begin{center} \funit=1.5pt                                       
\def\dot#1){\put#1){\circle*3}}
\def\negtri#1{\multiput(0.1,-0.05)(0.2,-0.1){#1}{{\funit=1pt \circle*{1}}}}
\savebox\link(0,0)[tl]{\bpic(1.8,0.9)\negtri8\epic}
\savebox\figc(60,20)[bl]{\bpic(60,20)
   \putlin(30,0,1,0,25) \putlin(15,20,3,-4,15) \putlin(0,0,1,0,30)
   \putvec(40,0,-1,0,0) \putvec(15,20,3,-4,9)  \putvec(15,0,-1,0,3)
   \multiput(14,21.3)(3.6,-1.8){11}{\usebox\link} 
   \lab(15,25)b2\dot(15,20) \lab(30,-5)t4\dot(30,0) \lab(0,-5)t3\dot(0,0)
   \lab(55,-5)t1\dot(55,0)  \lab(55,25)b5\dot(55,20) \epic}
\savebox\figb(60,20)[bl]{\bpic(60,20)
   \putlin(0,0,3,4,15) \putvec(0,0,3,4,9) \epic}
\savebox\figa(60,20)[bl]{\bpic(60,20)
   \putlin(55,0,0,1,20)\putvec(55,10,0,-1,3) \epic}
\bpic(195,35)\put(0,15){\usebox\figa}\put(0,15){\usebox\figb}
       \put(0,15){\usebox\figc} \lab(28,0)t{(A)}
   \put(140,15){\usebox\figb} \put(140,15){\usebox\figc} \lab(168,0)t{(B)}
\epic\\[6mm]
\savebox\figa(60,20)[bl]{\bpic(60,20)
   \putlin(30,0,1,0,25) \putlin(15,20,3,-4,15) 
   \putvec(40,0,-1,0,0) \putvec(15,20,3,-4,9)  
   \multiput(14,21.3)(3.6,-1.8){11}{\usebox\link} 
   \lab(15,25)b2\dot(15,20) \lab(30,-5)t4\dot(30,0) \lab(0,-5)t3\dot(0,0)
   \lab(55,-5)t1\dot(55,0)  \lab(55,25)b5\dot(55,20) \epic}
\bpic(195,60)\put(0,15){\usebox\figc} \lab(28,0)t{(C)}
   \put(140,15){\usebox\figa}  \put(140,15){\usebox\figb}  \lab(168,0)t{(D)}
\epic\\[9mm]
Fig. 2: Graphical representation of the models in table 2.
\end{center} \end{figure}

The first deviation from the above pattern occurs at Euler number 540.
Again all skeletons are unique, but this time the loop contains 3 fields.
The graphical representation is shown in fig.~2. The same graphs also
account for the remaining 3 spectra with $\c\ge450$, which are listed in
table~2.\footnote{The singlet spectrum $(278,53;-450)$
  with the smallest Euler number, on the other hand, is quite different:
  It comes from
  the configuration {\config 5 1 4 27 63 94 189 }, 	  
  which admits 5 different skeletons and does not require a loop. }
Here most of the entries are the same as in table~1. The only modification
is that the orbifold representation can now be obtained by cutting the
pointer at $X_2$ or the pointer at $X_3$, as is shown in figs.~2C and 2D,
respectively. This is indicated by a superscript of the orbifold
configuration; $b_I$
is the new exponent of $X_2$ or $X_3$ in the two cases. The model
with $\chi=512$ is the only one for which both pointers can be cut. Thus
we find 3 different orbifold representations, where $C\cap D$ indicates that
both pointers have been cut (see section~4 for more details). As before,
an additional pointer from $X_5$ to $X_1$ has to be supplemented in the
graphs 2C and 2D if the orbifold configuration descends from~2A. Each of
the models in table~2 can be realized in at least one invertible
configuration, but none of them do we find in a Fermat configuration.

For the spectrum with $\c=540$ we have two entries in table~2, because in
this case there is no obvious non-linear transformation relating the
following two polynomials:
\beq X_1^{44}X_4+X_2^{38}X_4+X_3^5X_2+X_4^3X_3+\e X_1^{30}X_2^{26}, \eql 4
\beq X_1^{52}X_4+X_2^{38}X_4+X_3^3X_2+X_4^5X_3+\e X_1^{30}X_2^{22}. \eql 5
The two configurations with 5 fields are of course related to
(\ref 4),~(\ref5) by the transformation (\ref z), with $n=2$ and a
relabelling of the fields. To search for a relation
between (\ref4) and (\ref5) I tried a general ansatz, allowing the
monomials of the skeleton of one of these polynomials to transform into an
arbitrary monomial of degree $d$ in the target configuration.
As there are 21 such monomials, this required the computation
of ${21!\017!}=143\,640$ determinants (which Mathematica did in 14 hours).
Unfortunately, none of these determinants is constant and non-vanishing.

As a simple check of the program I also used it to search for non-linear
symmetries of the above skeletons. It should have found at least the identity,
of course, but it also found an additional symmetry:
\beq X_i\to X_i\r^{c_i},\qqd \r=\({X_2^{19}\0X_1^{22}}\)^{1\0571},\qqd
     \{c_i\}=\{26,-30,6,-2\}, \eql a
where $p=571$ is the $105^{th}$ prime number. This transformation leaves
(\ref 4) invariant, whereas for (\ref 5) we have to use
$\r^{571}=X_2^{19}/X_1^{26}$ and $\{c_i\}=\{22,-30,10,-2\}$.
In both cases we have a non-linear $\ZZ_2$ symmetry of the skeleton.
Of course, as for all our non-linear transformations, we also have to
check that there exists a system of links that is consistent with the
transformation. In (\ref 4),~(\ref 5) the simplest invariant link that makes
the polynomial non-degenerate is already included. Surprisingly, most of the
monomials of degree $d$ that exist in these configurations are invariant
(the counting is done without trivial fields): For the
configuration of (\ref 4) there are 15 invariant monomials of degree $d$,
whereas the following three pairs of monomials are transposed by the
non-linear $\ZZ_2$ symmetry:
$(X_1^{44}X_4,X_2^{38}X_4)$, $(X_1^{52}X_2^7,X_1^8X_2^{45})$, and
$(X_1^{46}X_2^2X_3,X_1^2X_2^{40}X_3)$.   In the case of (\ref 5)
we have 18 invariant monomials and the two    transforming pairs
$(X_1^{52}X_4,X_2^{38}X_4)$ and $(X_1^{56}X_2^3,X_1^4X_2^{41})$.

For the time being we thus have to consider the \LG\ models based
on (\ref 4) and (\ref 5) as distinct. It cannot be excluded that there
exists some identification of the underlying conformal field theories,
for example as an orbifold with respect to the non-linear $\ZZ_2$ symmetry,
but I do not know how to check for this possibility.
Still, we can find some relation: As above, we can construct orbifold
representations of the models in the invertible configurations
$\IC_{(1,1,12,16)}[60]$ and $\IC_{(1,1,8,20)}[60]$ by cutting the loops.
We can then deform the potentials such that they become the transpose
of each other. Thus further orbifolding and a mirror map complete the
connection.

Let us finally note that almost all polynomials in table~1 have exponents 3
and 7, whereas those in table~2 generically have exponents 3 and 5.
This is not surprising, as the smallest values for the central
charge larger than  the critical value $c=3$ enter
the derivation of limits on the exponents in
non-degenerate potentials~\cite{cqf}.
The first of these 
come from the polynomials $X^3+Y^7$, $X^3+XY^5\sim X^5+XY^3$, $X^3+Y^8$,
and $X^3Y+XY^4$, 
with $c/3-1=1/21$, $1/15$, $1/12$, and $1/11$, 
respectively.\footnote{
Any accumulation point can only be approximated from below~\cite{cqf}. So
there must be an interval above $c=3$ with a finite spectrum, providing, in a
sense, a continuation of the ``exceptional'' cases $E_6$, $E_7$ and
$E_8$~\cite{agv}.}
It is then also not surprising that the models with an exponent~5
need an additional pointer. The presence of these particular exponents
is thus related to the requirement of a large Euler number
rather than to the absense of mirror symmetry.
What appears to be significant, then, is that
even in these exceptional cases there are plenty of non-linear relations,
and of their associated (hidden) symmetries.

\section{Another look at non-linear transformations}

It is straightforward to generalize the non-linear transformations encountered
above to arbitrary skeletons. With the only exception of the
non-linear symmetry (\ref a),
all of them had the effect of eliminating one pointer and changing one
exponent. So let us first consider the case of a pure loop and make the ansatz
\beq X_1^{a_1}X_2+ X_2^{a_2}X_3+\ldots+ X_n^{a_n}X_1=
     Y_1^{b_1}Y_2+ Y_2^{a_2}Y_3+\ldots+ Y_n^{a_n}, \eql L
with $Y_i=X_iX_1^{c_i}$. Then
$c_i=(-1)^{n-i}/(a_i\ldots a_n)$ for $i>1$ and $b_1=(a_1-c_2)/(1+c_1)$,
where $c_1$ is to be computed from the condition $\sum_{i=1}^n c_i=0$
for constant determinant. Because of our ansatz for the transformation,
this condition is equivalent to not changing the central charge.
The transformation makes sense if the exponent $b_1$, as computed from these
formulae, is integer.

Now we try to cut the pointer at $X_I$ in an unbranched tree of length $n$,
\beq X_1^{a_1}X_2+\ldots 
    +X_n^{a_n}   =   Y_1^{a_1}Y_2+\ldots
    Y_{I-1}^{a_{I-1}}~+~Y_I^{b_I}Y_{I+1}\ldots+Y_n^{a_n},\eql T
with the same ansatz for the transformation.
Here we find $c_i=(-1)^{I-i-1}/(a_i\ldots a_{I-1})$ for $i<I$ and $c_i=0$
for $i>I$. Again, $c_I$ has to be computed from the determinant condition
$\sum c_i=0$ and the new exponent $b_I=a_I/(1+c_I)$ should be integer.

For a given field $X_I$ in a general skeleton we can have a number of pointers
at $X_I$, which we take to be $X_1,\ldots,X_N$, and $X_I$ can point at
$X_{I+1}$ or can be of the Fermat type.
In the latter case we formally set $X_{I+1}=1$ and $c_{I+1}=0$.
Then the $X_I$-dependent part of the potential is
\beq \sum_{j=1}^N X_j^{a_j}X_I +X_I^{a_I}X_{I+1} =
     \sum_{j=1}^J Y_j^{a_j}+\sum_{j=J}^N Y_j^{a_j}Y_I+Y_I^{b_I}Y_{I+1},\eql B
where we try, without loss of generality, to cut the pointers from $Y_1,\ldots,
Y_J$ and to keep the remaining ones. From our ansatz we find
\beq b_I={a_I-c_{I+1}\01+c_I},~~~~~ c_j=\Biggl\{{~\hbox{$1/a_j$ ~~~~~ $j\le J$}
     \atop \hbox{$-c_I/a_j$ ~~~ $j>J$}}.\eeq
For any field $X_i\to X_k$ with a line of pointers connecting it to $X_I$
the corresponding $c_i$ is given recursively by $c_i=-c_k/a_i$.
If there is no such line of pointers, then $c_i=0$. Thus, in particular,
$c_{I+1}=0$ if $X_I$ is not in a loop. Eventually, $c_I$ has to be computed
from the condition that the sum of all $c_i$ has to vanish and the
new exponent $b_I$ should be integer.

With these formulae it is straightforward to check all possibilities to cut
a pointer in an arbitrary graph by a transformation of the form
\beq X_i\to X_I^{c_i}X_i,    \eql N
where $X_I$ is the target of that pointer. In fact, I computed
the orbifold part of tables~1 and~2 by writing a simple program, which does
just that (the entry ``Twist'' in the tables is given by the number $1+c_I$).
In the above sections we found only one transformation that is not an
iteration of (\ref N), namely the symmetry (\ref a),
but even that case is of the  form $X_i\to \r^{c_i}X_i$. Of course, in
case of branchings, one always has to make sure that the additional monomials
required for non-degeneracy can be chosen in a consistent way [for (\ref N)
it is sufficient to check the $X_I$-dependent links].

As non-linear transformations with constant determinant play such a prominent
role in \LG\ models, it would be important to have more checks of
the conjectured equivalence of the corresponding orbifolds. 
Such a check should be possible in the simplest case, namely with two
fields and a single pointer. Here the condition that the pointer can
be cut implies that $W=X^{(n-1)a}+XY^n$. The transformed potential $\5W=
{\5X}^{na}+{\5Y}^n$ is of the Fermat type and thus a tensor product of minimal
models (this case has already been studied in ref.~\cite{ls}).
As the transformation $X\to X^{n/(n-1)}$, $Y\to Y/X^{1/(n-1)}$ is $n$ to
$n-1$, the $\ZZ_n$ orbifold of the exactly solvable Fermat model should have
a hidden $\ZZ_{n-1}$ symmetry if the conformal field theories $W/\ZZ_{n-1}$
and $\5W/{\ZZ_n}$ are indeed identical.

It is easy to check that the charge degeneracies of the chiral ring agree in
the two cases, which both have central charge ${c/6}=1-(a+1)/(na)$~\cite{iv}:
In the pointer case the charges $q_i=n_i/d$ of the chiral fields are
$q_X={1/((n-1)a)}$ and $q_Y=(na-a-1)/(n(n-1)a)$. The projected untwisted sector
of the (c,c) ring of the orbifold is generated by
\beq \cR^u=\<\{Y^{n-1}\}\cup\{(XY)^jX^{(n-1)k}\;|\;
     j\le n-2,~ k\le a-1\}\>,                            \eql m
and in addition we have $n-2$ twisted (c,c) states, all of which have charge
$c/6$. The quantum $\ZZ_{n-1}$ symmetry, which goes with the
orbifolding~\cite{v}, only acts on these twisted states.
In the Fermat case, on the other hand, the charges are $q_{\5X}={1/na}$ and
$q_{\5Y}={1/n}$. The untwisted sector is generated by
\beq\5\cR^u=\<\{(\5X\5Y)^j\5X^{nk}\;|\;j\le n-2,~ k\le a-1\}\>,
          \eql n
and there are $n-1$ twisted chiral primary fields with non-trivial action
of the quantum $\ZZ_n$.
The charges of the monomials in (\ref m) and (\ref n) coincide as
functions of $j$ and $k$, but in the pointer case there is the additional
invariant contribution $Y^{n-1}$ to the chiral ring. This field has
indeed the right charge to replace one of the twisted states in the Fermat
case and should correspond to a linear combination of them.
We thus have some indications of how the hidden quantum $\ZZ_{n-1}$ symmetry
should act in the Fermat case, but we must leave the full investigation
for future work.

\section{Discussion}

We have analysed a number of models with large positive Euler
number for which no mirror is known.
All of them have surprisingly similar structures, which helped to find
a large class of non-linear transformations with constant determinants.
Although our models can be obtained from a number of different
configurations and orbifolds, the transformations could be used, in most
cases, to identify all of the apparently different representations. For
the only exception, the first spectrum in table~2, which has two apparently
different families of representations, we instead found a non-linear $\ZZ_2$
symmetry.
Another rule with one exception (the last spectrum in table~1) is
that each of the models can be obtained as an orbifold in an invertible
configuration. For a deformation, which however has different
symmetries, we would thus know how to construct the mirror.

Non-linear transformations could of course be used to reduce considerably
the number of polynomials that have
to be investigated, for example, in the classification of orbifolds.
With the present state of technology this would, however, be of limited
value, as we would lose many symmetries that are manifest only in the
redundant configurations.
Further studies are therefore in order: It would be important to find
out whether the related orbifolds indeed correspond to identical
conformal field theories. And if this is the case, we would need
a technology for exploiting the implied hidden and non-linear symmetries.

It may be hoped that further investigation of our set of models will also be
of help in the search for their mirror partners. But so far, unfortunately,
the question remains: Where are the mirror manifolds?

{\it Note added:}
After submitting the present work for publication I was kindly informed by
M.R. Plesser about ref.~[21], where related ideas are pursued.
That paper, in particular, clearifies the meaning of non-linear field
transfromations with constant Jacobian in the context of toric geometry.
Furthermore, it is argued that the corresponding field theories flow to
fixed points with the same complex structure, but different values of the
K"ahler moduli. For a number of examples this is confirmed by identifying
the implied classical symmetries of the mirror manifolds.

{\it Acknowledgements.} It is a pleasure to thank Per Berglund,
Philip Candelas and J"urgen Fuchs for helpful discussions, Jan Louis for
triggering the present investigation, and in particular Harald Skarke for
the enjoyable collaboration on the results of which this work is based.


\end{document}